%% file: main.tex
\documentclass[conference]{IEEEtran}
\usepackage[utf8]{inputenc}
\usepackage{hyperref}

\ifCLASSINFOpdf
\else
\fi
\usepackage{amsmath,amssymb}
\usepackage{mathtools}
\usepackage{tikz}
\usetikzlibrary{arrows,fit,positioning}
\usepackage{pgfplots}
\pgfplotsset{width=10cm,compat=1.9}
\usepackage{tabularx}
\usepackage{colortbl}
\usepackage{multirow}
\usepackage{arydshln}
\usepackage[caption=false,font=footnotesize]{subfig}
\usepackage[export]{adjustbox}
\usepackage{soul}

\newcommand\lsub[1]{
    \prescript{}{#1}
}

\newcommand{\mycc}{\cellcolor{lightgray}}

\begin{document}

\title{Automatic Datapath Optimization using E-Graphs}

\author{\IEEEauthorblockN{Samuel Coward}
\IEEEauthorblockA{Numerical Hardware Group\\
Intel Corporation\\
Email: samuel.coward@intel.com}
\and
\IEEEauthorblockN{George A.~Constantinides}
\IEEEauthorblockA{Electrical and Electronic Engineering\\
Imperial College London\\
Email: g.constantinides@imperial.ac.uk}
\and
\IEEEauthorblockN{Theo Drane}
\IEEEauthorblockA{Numerical Hardware Group\\
Intel Corporation\\
Email: theo.drane@intel.com}}

% \author{\IEEEauthorblockN{Anonymous}
% \IEEEauthorblockA{Anonymous\\
% Anonymous\\
% Email: Anonymous}

%Telephone: (800) 555--1212, Fax: (888) 555--1212}


\maketitle

\begin{abstract}
Manual optimization of Register Transfer Level (RTL) datapath is commonplace in industry but holds back development as it can be very time consuming. We utilize the fact that a complex transformation of one RTL into another equivalent RTL can be broken down into a sequence of smaller, localized transformations. By representing RTL as a graph and deploying modern graph rewriting techniques we can automate the circuit design space exploration, allowing us to discover functionally equivalent but optimized architectures. We demonstrate that modern rewriting frameworks can adequately capture a wide variety of complex optimizations performed by human designers on bit-vector manipulating code, including significant error-prone subtleties regarding the validity of transformations under complex interactions of bitwidths. The proposed automated optimization approach is able to reproduce the results of typical industrial manual optimization, resulting in a reduction in circuit area by up to 71\%. Not only does our tool discover optimized RTL, but also correctly identifies that the optimal architecture to implement a given arithmetic expression can depend on the width of the operands, thus producing a library of optimized designs rather than the single design point typically generated by manual optimization. In addition, we demonstrate that prior academic work on maximally exploiting carry-save representation and on multiple constant multiplication are both generalized and extended, falling out as special cases of this paper.

\end{abstract}


% \fi
\IEEEpeerreviewmaketitle

%%%%%%%%%%%%%%%%%%%%%%%%%%%%%%%%%%%%%%%%%%%%%%%%%%%%%%%%%%%
% INTRODUCTION
%%%%%%%%%%%%%%%%%%%%%%%%%%%%%%%%%%%%%%%%%%%%%%%%%%%%%%%%%%%
\section{Introduction}\label{sect:intro}
In industry and academia, Register Transfer Level (RTL) development is limited to minimal design space exploration due to the complexity of design space and is slowed down by long debug timelines. RTL optimization of datapath designs is still often a manual task, as synthesis tools are unable to achieve the results of a skilled engineer \cite{Synopsys2019CodingSynthesis}. %The ever-growing complexity of digital circuit designs makes automatic optimization a constantly changing challenge. Automated and formally verified optimization can reduce the frequency with which bugs are introduced improving production timelines. \gc{Preceding two sentences are fine, but can be cut if looking for space.}

A key observation is that manual RTL optimization is typically performed by applying a number of known `useful' transformations to a design. These transformations, and their domain of validity, are accumulated through years of engineer design experience. In combination, these transformations may result in substantial changes to the underlying RTL. Apart 
from some simple transformations implemented automatically in modern ASIC design tools \cite{Synopsys2021DesignS-2021.06-SP2}, the process of determining a sequence of transformations to apply to an RTL design is currently based on designer intuition \cite{Verma2009ChallengesCircuits},
largely due to the non-convex nature of the design space: it is often necessary to apply an early transformation
that results in a worse-quality circuit before then applying a later one leading to an overall improvement. It is this process we seek to automate.

Such automation facilitates the creation of bitwidth dependent architectures, where different parameterisations may result in different architectures: the best design approach for narrow bitwidths may not be the best design approach for wider bitwidths. A range of RTLs automatically generated from a single parameterisable input retains the ease-of-use benefits of parameterisable RTL without sacrificing quality.

Existing commercial synthesis tools are capable of merging together consecutive additions in a circuit design to make best use of carry-save representations \cite{Synopsys2019CodingSynthesis}. However when arithmetic is interspersed with logic, the tools frequently miss potential optimization opportunities \cite{dataflow2008verma}.

We aim to leverage existing commercial synthesis tools by transforming RTL to a form the existing tools can maximally optimize. We focus on combinational RTL, although the techniques described are equally applicable to pipelined designs via retiming. Given a design in the form of an RTL implementation $R$, we aim to find an RTL implementation $R'$ that minimises $\textrm{cost}(R')$ for some cost function, such that the two RTLs are functionally equivalent, $R\simeq R'$. We define the equivalence relation $\simeq$\ as $R\simeq R'$ if and only if for all possible inputs, all outputs of $R$ and $R'$ are equal.

We represent such RTL as a data-flow graph, where operators and operands are represented by nodes with edges, labelled by bitwidth, connecting operands and operators. This graphical representation allows us to formulate the problem as a graph optimization problem, where we are allowed to manipulate the graph with equivalence-preserving transformations. This formulation allows us to take advantage of recent advances in e(quivalence)-graph and equality saturation~\cite{Willsey2021Egg:Saturation} technology, discussed in Section \ref{sec:background}, alongside previous motivating work in automated RTL optimization and design. Application of e-graphs to the RTL optimization problem is presented in Section \ref{sec:methodology}. We demonstrate results in Section \ref{sect:results} and validate
our cost metric in Section \ref{sect:cost_validation}. 

The paper contains the following novel contributions:
\begin{itemize}
    \item application of e-graphs and equality saturation to automate datapath RTL optimization,
    \item a precisely defined set of rewrites that facilitate efficient design space exploration together with their domains of applicability in designs utilizing multiple bitwidths,
    \item an automated method to optimize architectures as a function of bitwidth parameters,
    \item quantification of a `noise floor' in datapath logic synthesis using `fuzzing' techniques from software testing.
\end{itemize}

%%%%%%%%%%%%%%%%%%%%%%%%%%%%%%%%%%%%%%%%%%%%%%%%%%%%%%%%%%%
% BACKGROUND
%%%%%%%%%%%%%%%%%%%%%%%%%%%%%%%%%%%%%%%%%%%%%%%%%%%%%%%%%%%
\section{Background} \label{sec:background}
\subsection{Datapath Optimization} \label{subsec:datapath_opt}
A useful example of transformation-based datapath improvement comes from Verma, Brisk and Ienne \cite{dataflow2008verma}, which automatically applies data-flow transformations to maximally exploit carry-save representation. Their primary objective is to cluster additions together in the data-flow graph, a useful target as full carry-propagate addition is generally expensive and can often be avoided. This can be done by deploying compressor trees, circuits taking three or more input words which get reduced to two output words: a carry and a save. Using a carry-propagate adder to sum the carry and the save returns the sum of all the inputs. This can be beneficial as, for example, combining two consecutive carry-propagate adders into a compressor tree and one carry-propagate adder saves the cost of one carry-propagate adder at the expense of a compressor tree. 
%The simplest implementation is a three-input carry-save adder (CSA). These simple CSAs can be used as building blocks to make more complex compressor trees which take more inputs. 
We generalize this work using our methodology, which is able to replicate the results obtained in \cite{dataflow2008verma} as a special case. 

Another well-studied transformation beyond the reach of standard commercial synthesis tools is the multiple constant multiplication (MCM) problem \cite{Gustafsson2007AProblems,Hartley1996SubexpressionMultipliers}. The MCM problem asks, given a set of integer coefficients $\{a_1, ..., a_n\}$ and variable $x$, what is the optimal architecture to compute the set $\{a_1\times x, ..., a_n \times x\}$? Competing solutions use a fixed number representation of the constants \cite{Hartley1996SubexpressionMultipliers}, often canonical signed digit (CSD) representation \cite{Ercegovac2004DigitalArithmetic}, and/or deploy an adder graph algorithm \cite{Gustafsson2007AProblems}. A transformation based approach also captures both of these methods.

In addition to these special cases, there is a wide variety of transformations that can be captured through standard arithmetic rewrites, {\em e.g.}~associativity, distributivity, {\em etc.} Often these rewrites interact with each other, in the sense that applying one type of transformation opens or closes the door to applying a different class of transformation.

\subsection{E-graphs and Equality Saturation}
\label{subsec:equality_saturation}
\begin{figure}
    \centering
    \subfloat[Initial e-graph contains \newline $(2\times x)>>1$] {\includegraphics[scale=0.4]{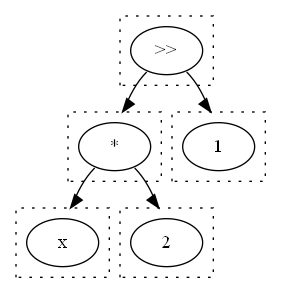}}
    % \qquad
    \subfloat[Apply $x\times 2 \rightarrow x<<1$] {\includegraphics[scale=0.4]{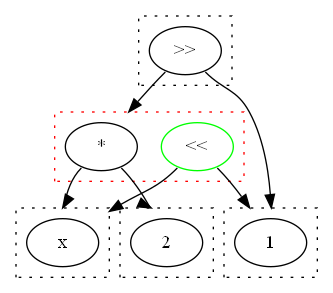}}
    \qquad
    \subfloat[Apply $(x<<s)>>s \rightarrow x$] {\includegraphics[scale=0.4]{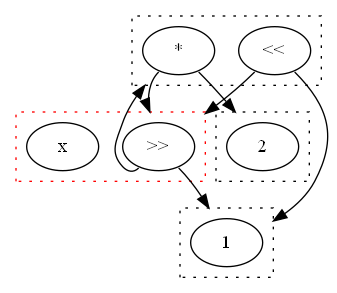}}
    \caption{E-graph rewriting for standard integer arithmetic. Dashed boxes represent e-classes of equivalent expressions. Green nodes represent newly added nodes. Red dashed boxes highlight which e-class has been modified.
    }
    \label{fig: e-graph_example}
\end{figure}

Equivalence graphs, commonly called e-graphs, provide a dense representation of equivalence classes (e-classes) of expressions \cite{Nelson1980TechniquesVerification}. Often found in theorem provers, this data structure enables a graph optimization technique called equality saturation \cite{Willsey2021Egg:Saturation,Joshi2002Denali:Superoptimizer,Tate2009EqualityOptimization}. The e-graph represents expressions, where the nodes, known as e-nodes, represent functions (including variables and constants, as 0-arity functions) and are partitioned into a set of e-classes. The intuition is that e-classes can be used to compactly represent {\em equivalent} expressions, whose evaluation always leads to the same result. Edges represent function inputs and are from e-nodes to e-classes; see Figure \ref{fig: e-graph_example}, where dashed lines represent e-class boundaries, solid ellipses are e-nodes and arrows are edges. We define $\mathcal{C}$ to be the set of e-classes, $\mathcal{N}$ the set of e-nodes and $E \subseteq \mathcal{N} \times \mathcal{C}$ the set of edges. We also introduce $\mathcal{N}_c$ to denote the set of e-nodes in a given e-class $c$.  

Rewrites define equivalences over expressions, for example $x+x \rightarrow 2 \times x$ says that $x+x$ is equivalent to $2\times x$. Such rewrites are applied constructively to the e-graph, meaning that the left hand side of the rewrite remains in the data structure. Constructive rewrite application avoids the concern of which order to apply rewrites in. As rewrites are applied, the e-graph grows monotonically, representing more and more equivalent expressions, and hence naturally capturing the interaction between different rewrite rules.

Equality saturation provides us with a stopping condition. At the point where further rewrites add no additional information, we say that the e-graph has saturated. From an e-graph representing potentially infinitely many equivalent expressions we may choose the ``best'' expression \cite{Willsey2021Egg:Saturation}.

\texttt{egg} is a recent Rust e-graph library, which is intended to be a general purpose and reusable implementation \cite{Willsey2021Egg:Saturation}. It adds powerful performance optimizations over existing, usually bespoke, e-graph implementations along with some useful additional features. It has been used to automatically improve the numerical stability of floating point expressions~\cite{Panchekha2015AutomaticallyExpressions}, map programs onto hardware accelerators \cite{Smith2021PurePearl} and optimize linear algebra \cite{Wang2020SPORES:Algebra}. To build a functioning e-graph optimization tool, \texttt{egg} must be supplied with a language definition -- that is a set of operator names together with their arity, and a rewrite set -- that is a set of equivalences over the given language definition.
%\begin{itemize}
%    \item Language definition -- set of operator names together with their arity
%    \item Rewrite set -- rewrites that define equivalences over the given language definition
%\end{itemize}

% Removed this as we use ILP extraction
%\gc{Is the following paragraph needed if we don't use this functionality?}
%To select the best expression from the final e-graph, \texttt{egg} uses a process called extraction. It traverses the e-graph working bottom-up towards the root, using the supplied local cost function to choose the best equivalent node from each equivalence class. The cost function is local, as the node cost is only a function of the node itself and its children. The best expression is then given by the node with the lowest cost that is found in the same equivalence class as the original root of the initial e-graph. 

%%%%%%%%%%%%%%%%%%%%%%%%%%%%%%%%%%%%%%%%%%%%%%%%%%%%%%%%%%%
% METHODOLOGY
%%%%%%%%%%%%%%%%%%%%%%%%%%%%%%%%%%%%%%%%%%%%%%%%%%%%%%%%%%%
\section{Methodology}\label{sec:methodology}
This section demonstrates how e-graphs can be applied to the RTL optimization problem. We use a natural graphical representation of RTL, using data-flow graphs allowing us to fit the optimization problem into the e-graph framework. In this framework the e-graph contains classes of equivalent bitvector manipulating expressions and the rewrites transform such expressions to alternative equivalent expressions. The tool parses input Verilog using Yosys \cite{Wolf2013Yosys-ASuite} and converts it into nested S-expressions in Common Lisp \cite{Steele1990CommonLanguage}. 

\texttt{term::=(operator [term] [term]\ldots [term])}

\begin{figure}
    \centering
    \input{flow_diagram}
    \caption{Flow diagram describing the operation of the tool.}
    \label{fig:flow_diagram}
\end{figure}
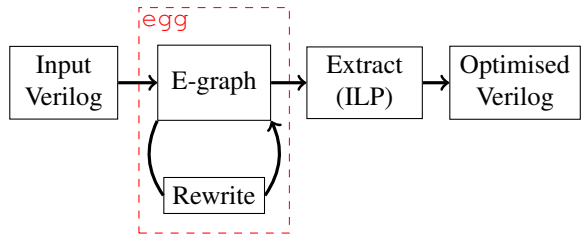

\noindent The syntax is defined by the language described in Section \ref{subsect:lang}. These expressions are converted into e-graphs by \texttt{egg}. From the e-graph we extract a nested S-expression from which RTL is automatically generated, writing one operation per line. Figure~\ref{fig:flow_diagram} provides a flow diagram of the tool.

\begin{table}
    \centering
    \caption{Operators defined in our \texttt{egg} implementation of RTL optimization. We include the architecture chosen for theoretical cost assignment.}
    \begin{tabular}{c|c|c|c}
        Operator                         & Symbol & Arity & Architecture \\
        \hline            
        Left/Right Shift                 & $<</>>$&  2    & Mux Tree \\
        Addition/Subtraction             & +/-&  2    & Prefix Adder (PA) \cite{Beaumont-Smith2001ParallelDesign}\\ 
        Negation                         & - &  1    & PA\\
        Multiplication                   & $\times$ &  2    & Booth Radix-4 \cite{Koren2018ComputerAlgorithms} \\
        Multiplexer                      &$\cdot ? \cdot : \cdot$&  3    & Mux gates \\
        Not/Inversion                    & $\sim$ &  1    & One-input gates\\
        Concatenate                      & \{,\} &  n    & Wiring\\
        Comparison                       &$>/<$&  2    & PA\\
        \cdashline{1-4}
        Sum                              & \texttt{SUM} &  n    & CSA and PA\\
        Muxed Mult Array                        & \texttt{MUXAR}&  3    & Array Reduction and PA \\
        Fused Multiply-Add               & \texttt{FMA} & 3 & Booth Radix-4 \\
        %Union (UNION)        & \\
    \end{tabular}
    \label{tab:operator_arch}
\end{table}
\subsection{Language}\label{subsect:lang}
We consider the problem from the abstraction level of a Verilog parser and operate on finite length bitvectors. We target bitvector arithmetic and bitwise Boolean operations as they form the basis of low-level datapath optimization. Including the bitwidth of these vectors is crucial to correctly model the circuit's behaviour. Bitwidths are also essential to correctly evaluate the cost of a given operation, clearly an 8-bit addition is less expensive than a 32-bit addition. The defined operations represented by nodes in the e-graph are described in Table \ref{tab:operator_arch}, and are defined for all inputs. 

\begin{figure}
    \centering
    % \subfloat[Consecutive additions]{\includegraphics[width=.35\columnwidth]{two_adds.png}}
    \subfloat[Consecutive additions]{\includegraphics[scale=.45]{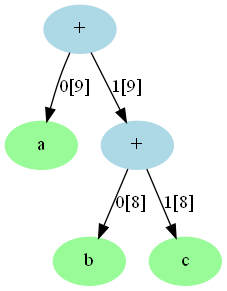}}
    \qquad
    % \subfloat[Merged additions encoded as a \texttt{SUM}] {\includegraphics[width=.55\columnwidth]{two_sum.png}
    \subfloat[Merged additions encoded as a \texttt{SUM}] {\includegraphics[scale=.45]{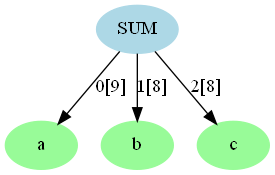}
    \label{fig: merged_adds}}
    \caption{The edge labels contain the operand's index and the operand's bitwidth in square brackets.}
    \label{fig: sum_node}
\end{figure}

The first set in the table is a subset of the operators defined in Verilog \cite{Thomas2008TheLanguage}. These operators are fundamental in most arithmetic circuit designs and manipulating designs using them can have significant effects on power, performance and area. Bitwidth information is included in the e-graph as edge labels between the nodes. In addition to these basic operators, we introduce a second set of operators in 
Table~\ref{tab:operator_arch} (below the dashed line), which encode the merging and sharing capabilities of modern synthesis tools. In \cite{dataflow2008verma}, the benefits of merging adjacent additions to maximise the use of carry-save are exploited. We are able to capture the same effect in an automated manner through introduction of a \texttt{SUM} node in our language definition, which combines an arbitrary number of additions into a single node. This node encodes the compressor tree and carry-propagate adder described in Section \ref{sect:intro}. Figure~\ref{fig: sum_node} shows how consecutive additions can be rewritten as a \texttt{SUM} node. Other such merging operators are the fused multiply-add (\texttt{FMA}) and the less familiar muxed multiplication array (\texttt{MUXAR}), which blends two disjoint multiplication arrays into one. These are discussed further in Section \ref{subsect:rewrites}.

%In Verilog the bitwidth of an expression is either context-determined (meaning the left-hand side of an assignment ($=$) determines the width of the right-hand side), or self-determined (meaning the expression width is determined only by operand widths). Synopsys' Coding Guidelines recommend avoiding self-determined expressions to avoid ambiguity \cite{Synopsys2019CodingSynthesis}, therefore we assume all expressions are context-determined. An edge labelling provides bitwidths for all inputs and outputs of an operator, allowing us to capture context-determined behaviour.  
%\gc{How important is the previous bitdwidth / Verilog semantics discussion in {\em this} paper, given that the translation process is manual? Should we leave this for a version of this work where we do the translation automatically?}

\subsection{Rewrites} \label{subsect:rewrites}

\begin{table*}
    \centering
    \caption{The set of bitwidth dependent rewrites supplied to \texttt{egg}. Left subscript notation, $\lsub{p}x$, denotes a bitvector $x$ with length $p$ bits. The $*$ operation represents both $\{+,\times\}$. The square brackets represent Verilog bit slicing, where $a[x:y]$ means we take bits $x$ down to $y$ of $a$. The rules are conditionally applied according to column 4, which is sufficient but not always necessary.}
    \begin{tabular}{|c|c|c|c|}
    \hline
    
    Class & Name & Left-hand Side $\rightarrow$ Right-hand Side & Sufficient Condition\\
    \hline
    
    \multirow{13}{5em}{Bitvector Arithmetic Identities} & Commutativity   & $\lsub{r}(\lsub{p}a * \lsub{q}b)\rightarrow\lsub{r}(\lsub{q}b * \lsub{p}a)$ & True\\
    %----------------------------------------------------------------------------------------------------------------------------------------------
    
    &\mycc     &\mycc    &\mycc   $(q\geq t \lor r+s\leq q)$\\
    &\multirow{-2}{7em}{\mycc \centering Mult Associativity}  &\mycc  \multirow{-2}{18em}{\centering $\lsub{t}(\lsub{u}(\lsub{p}a \times \lsub{r}b) \times \lsub{s}c) \rightarrow \lsub{t}(\lsub{p}a \times \lsub{q}(\lsub{r}b \times \lsub{s}c))$}&\mycc   $\land (u\geq t \lor p+r \leq u)$\\
    %----------------------------------------------------------------------------------------------------------------------------------------------
    & \multirow{2}{7em}{\centering Add Associativity}   & \multirow{2}{18em}{$\lsub{t}(\lsub{u}(\lsub{p}a + \lsub{r}b) + \lsub{s}c) \rightarrow \lsub{t}(\lsub{p}a + \lsub{q}(\lsub{r}b + \lsub{s}c))$} & $(q\geq t \lor \max(r,s) < q)$\\
    & & & $\land (u\geq t \lor \max(p,r) < u)$\\
    %----------------------------------------------------------------------------------------------------------------------------------------------
    &\mycc   &\mycc  &\mycc  \\
    &\multirow{-2}{7em}{\mycc \centering Distribute Mult over Add} &\mycc  \multirow{-2}{25em}{\centering $\lsub{r}(\lsub{p}a \times \lsub{q}(\lsub{s}b + \lsub{t}c)) \rightarrow \lsub{r}(\lsub{u}(\lsub{p}a \times \lsub{s}b) + \lsub{v}(\lsub{p}a \times \lsub{t}c))$}&\mycc  \multirow{-2}{10em}{\centering $\min(q,u,v) \geq r$}\\
    %----------------------------------------------------------------------------------------------------------------------------------------------
    &  Sum Same        & $\lsub{q}(\lsub{p}a + \lsub{p}a) \rightarrow \lsub{q}(\lsub{2}2 \times \lsub{p}a)$ & True\\
    %----------------------------------------------------------------------------------------------------------------------------------------------
    &\mycc  Mult Sum Same   &\mycc  $\lsub{r}(\lsub{s}(\lsub{p}a \times \lsub{q}b) + \lsub{q}b)\rightarrow\lsub{r}(\lsub{t}(\lsub{p}a + \lsub{1}1) \times \lsub{q}b)$ &\mycc  $t>p \land s\geq p+q$ \\
    %----------------------------------------------------------------------------------------------------------------------------------------------
    & Add Zero & $\lsub{p}(\lsub{p}a + \lsub{q}b)\rightarrow\lsub{p}(a)$ & $b \equiv 0 \mod 2^p$ \\
    %----------------------------------------------------------------------------------------------------------------------------------------------
    &\mycc  Sub to Neg      &\mycc  $\lsub{r}(\lsub{p}a - \lsub{q}b)\rightarrow\lsub{r}(\lsub{p}a + \lsub{q}(-\lsub{q}b))$ &\mycc  True\\
    %----------------------------------------------------------------------------------------------------------------------------------------------
    &  Mult by One     & $\lsub{p}(\lsub{p}a \times \lsub{q}b)\rightarrow\lsub{p}(a)$ & $b \equiv 1 \mod 2^p$\\
    %----------------------------------------------------------------------------------------------------------------------------------------------
    &\mycc  Mult by Two     &\mycc  $\lsub{r}(\lsub{p}a \times \lsub{2}2)\rightarrow\lsub{r}(\lsub{p}a << \lsub{1}1)$ &\mycc  True\\
    \hline
    
    \multirow{6}{5em}{Bitvector Logic Identities} & Merge Left Shift & $\lsub{r}(\lsub{u}(\lsub{p}a<<\lsub{q}b)<<\lsub{s}c)\rightarrow\lsub{r}(\lsub{p}a << \lsub{t}(\lsub{q}b+\lsub{s}c))$ & $t>\max(q,s) \land u \geq r$\\
    %----------------------------------------------------------------------------------------------------------------------------------------------
    &\mycc  Merge Right Shift &\mycc  $\lsub{r}(\lsub{u}(\lsub{p}a>> \lsub{q}b) >> \lsub{s}c)\rightarrow\lsub{r}(\lsub{p}a >> \lsub{t}(\lsub{q}b+\lsub{s}c))$ &\mycc  $t>\max(q,s) \land u\geq p$ \\
    %----------------------------------------------------------------------------------------------------------------------------------------------
    & Redundant Sel     & $\lsub{p}(\lsub{1}b ? \lsub{p}a : \lsub{p}a) \rightarrow \lsub{p}a$ & True\\
    %----------------------------------------------------------------------------------------------------------------------------------------------
    &\mycc  Neg Not           &\mycc  $\lsub{r}(- \lsub{p}a)\rightarrow\lsub{r}(\lsub{p}(\sim(\lsub{p}a)) + \lsub{1}1)$ &\mycc  $r\leq p$\\
    %----------------------------------------------------------------------------------------------------------------------------------------------
    %& Not Neg           & $\lsub{r}(\lsub{p}(\sim(\lsub{p}a)) + \lsub{1}1)\rightarrow\lsub{r}(- \lsub{p}a)$ & $r\leq p$\\
    %----------------------------------------------------------------------------------------------------------------------------------------------
    & Not over Con      &  $\lsub{r}(\sim(\lsub{q+s}\{\lsub{q}a,\lsub{s}b\}))\rightarrow\lsub{r}\{\lsub{q}(\sim(\lsub{q}a)),\lsub{s}(\sim(\lsub{s}b))\}$ &  $q+s \geq r$\\
    \hline
    
    \multirow{3}{5em}{Constant Expansion} & \mycc & \mycc $\lsub{r}(\lsub{q}c \times \lsub{p}x)\rightarrow $ & \mycc \\
    & \multirow{-2}{7em}{\mycc  \centering Mult Constant} & \mycc $\lsub{r}(\lsub{r}(\lsub{q}(\lsub{2}2\times \lsub{q-1}c[q-1:1]) \times \lsub{p}x) + \lsub{p}(\lsub{1}c[0] \times \lsub{p}x))$ &  \mycc \multirow{-2}{7em}{\centering $c$ constant}  \\
    %----------------------------------------------------------------------------------------------------------------------------------------------
    &  One to Two Mult &  $\lsub{p}(\lsub{1}1\times \lsub{p}x)\rightarrow\lsub{p}(\lsub{q}(\lsub{2}2\times \lsub{p}x) - \lsub{p}x)$ &  $q > p$\\
    \hline
    
    \multirow{8}{5em}{Arithmetic Logic Exchange} & \mycc Left Shift Add  & \mycc $\lsub{r}(\lsub{s}(\lsub{p}a + \lsub{q}b) << \lsub{t}c) \rightarrow \lsub{r}(\lsub{u}(\lsub{p}a << \lsub{t}c) + \lsub{u}(\lsub{q}b << \lsub{t}c))$ & \mycc $(s\geq r \lor \max(p,q)<s) \land u\geq r$ \\
    %----------------------------------------------------------------------------------------------------------------------------------------------
    &   &   &  $q\geq t \land s\geq p + 2^u - 1 $\\
    & \multirow{-2}{7em}{ \centering Add Right Shift}& \multirow{-2}{28em}{ $\lsub{r}(\lsub{p}a + \lsub{q}(\lsub{t}b >> \lsub{u}c))\rightarrow\lsub{r}(\lsub{v}(\lsub{s}(\lsub{p}a << \lsub{u}c) + \lsub{t}b) >> \lsub{u}c)$}&  $  \land v > \max(s,t)$\\
    %----------------------------------------------------------------------------------------------------------------------------------------------
    & \mycc Left Shift Mult   & \mycc $\lsub{r}(\lsub{t}(\lsub{p}a\times \lsub{q}b)<<\lsub{u}c)\rightarrow\lsub{r}(\lsub{v}(\lsub{p}a<<\lsub{u}c)\times \lsub{q}b)$  & \mycc $t\geq r \land v\geq r$\\ 
    %----------------------------------------------------------------------------------------------------------------------------------------------
    &  \multirow{2}{7em}{\centering Sel Add}          &  $\lsub{r}(\lsub{1}e ? \lsub{r}(\lsub{p}a + \lsub{q}b) : \lsub{r}(\lsub{p}c + \lsub{q}d))\rightarrow$ &  \multirow{2}{4em}{\centering True} \\
    &         &  $\lsub{r}(\lsub{p}(\lsub{1}e ? \lsub{p}a : \lsub{p}c) + \lsub{q}(\lsub{1}e ? \lsub{q}b : \lsub{q}d))$ &  \\
    %----------------------------------------------------------------------------------------------------------------------------------------------
    &\mycc  Sel Add Zero      &\mycc  $\lsub{p}(\lsub{1}e ? \lsub{p}(\lsub{p}a + \lsub{q}b) : \lsub{p}c)\rightarrow\lsub{p}(\lsub{p}(\lsub{1}e ? \lsub{p}a : \lsub{p}c) + \lsub{q}(\lsub{1}e ? \lsub{q}b :  \lsub{q}0))$ & \mycc True\\
    %----------------------------------------------------------------------------------------------------------------------------------------------
    &  Move Sel Zero     &  $\lsub{r}(\lsub{p}(\lsub{1}b ? \lsub{p}0 : \lsub{p}a) \times \lsub{q}c)\rightarrow\lsub{r}(\lsub{p}a \times \lsub{q}(\lsub{1}b ? \lsub{q}0 : \lsub{q}c))$ &  True\\
    %----------------------------------------------------------------------------------------------------------------------------------------------
    &\mycc  Concat to Add &\mycc  $\lsub{r}\{\lsub{p}a,\lsub{q}b\}\rightarrow\lsub{r}(\lsub{s}(\lsub{p}a<<\lsub{u}q) + \lsub{q}b)$ &\mycc  $s\geq p + 2^u - 1 \land u \geq \lceil \log_2(q+1) \rceil$\\
    \hline
    
    \multirow{4}{5em}{Merging Ops} &  &  $\lsub{q_1}(\lsub{p_1}a1 + \lsub{q_2}(\lsub{p_2}a2 + \lsub{q_3}(\lsub{p_3}a3+...+\lsub{p_n}an)...))\rightarrow$ &  $q_i > \max(p_i, q_{i+1}), i=1,...,n-2$\\
    &  \multirow{-2}{7em}{\centering Merge Additions} &  $\lsub{q_1}(\texttt{SUM}(\lsub{p_1}a1, \lsub{p_2}a2,..., \lsub{p_n}an))$ &  $\land q_{n-1} > \max(p_{n-1}, p_n)$\\
    %----------------------------------------------------------------------------------------------------------------------------------------------
    &\mycc        &\mycc  $\lsub{t}(\lsub{s}(\lsub{q}a \times \lsub{r}b) + \lsub{s}(\lsub{q}c \times \lsub{r}(\sim(\lsub{r}b))))$ &\mycc  \\
    &  \multirow{-2}{7em}{\mycc \centering Merge Mult Array}  &\mycc   $\rightarrow\lsub{t}(\texttt{MUXAR}(\lsub{r}b, \lsub{q}a, \lsub{q}c))$ &\mycc \multirow{-2}{7em}{$s\geq q+r \land t>s$} \\
    %----------------------------------------------------------------------------------------------------------------------------------------------
    &  FMA Merge      &  $\lsub{t}(\lsub{s}(\lsub{p}a \times \lsub{q}b) + \lsub{r}c)\rightarrow\lsub{t}(\texttt{FMA}(\lsub{p}a, \lsub{q}b, \lsub{r}c))$ & $s\geq p+q \land t> \max(s,r)$ \\
    \hline
    \end{tabular}
    \label{tab:rewrites}
\end{table*}

We have identified and formalized rules that capture many of the manual transformations that Intel's Numerical Hardware Group regularly deploy manually. The set of rewrites described in Table \ref{tab:rewrites} define equivalences over expressions. Each of the rewrites incorporates bitwidth information, where bitwidths are associated with operands. We introduce a left subscript notation, $\prescript{}{p}{x}$, to denote a bitvector $x$ with length $p$ bits. Left subscript takes the lowest operator precedence. 

\texttt{egg} supports conditional rewrites. We make use of this facility to allow us to decide whether a rule may be applied based on its bitwidth information. The sufficient (but not always necessary) conditions under which each rule applies are described in Table \ref{tab:rewrites}. As an example, consider the ``Distribute Mult Over Add'' rewrite, where the sufficient condition is a function of the bitwidth parameters, $\min(q,u,v)\geq r$. When this condition is satisfied the rewrite can be safely applied as the equivalence holds. For some rewrites, a parameter appears on the right hand side but not on the left, for example $q$ in the associativity rewrites. This corresponds to a condition constraining but not fully determining the values that these undefined parameters can take, in which case we select the minimum feasible bitwidth. The conditions have been validated by creating a parameterizable SMT query for each rewrite using the theory of fixed size bitvectors \cite{Barrett2016TheSMT-LIB}, which have been checked for all combinations of bitwidths in \{1,...,10\}, however a formal equivalence check between the generated and original RTL ensures that we do not need to trust the correctness of our rewrites or of the \texttt{egg} library. 

Arithmetic Logic Exchange rewrites are inspired by a set taken from \cite{dataflow2008verma}, which focus on the interplay of logic and addition. To extend the prior work we have added rewrites that are able to move other arithmetic operators and we have generalized the rules so they are able to be applied in a multiple bitwidth setting.

The ``Merging Ops'' rewrites allow us to correctly evaluate the cost of specific sequences of operations, that logic synthesis tools are able to effectively optimize. We have seen that the \texttt{SUM} node merges consecutive additions, highlighting that the area usage for consecutive additions is not additive. This captures what logic synthesis tools will do to an expression such as $a+b+c$, converting it to a compressor tree and deploying a single carry-propagate adder. Further merging optimization capabilities of Synopsys Design Compiler are documented \cite{Synopsys2019CodingSynthesis}. The ``Merge Mult Array'' rewrite does not make use of carry-save format but identifies that two disjoint multiplication arrays can be merged. 
%For a $q$ bit operand $a$, where $a[i]$ represents bit $i$ of $a$ and $t$ is large enough to store $(p-1)$ to full precision.
Letting $a[i]$ represent bit $i$ of $a$ and $u = \lceil\log_2(r) \rceil$, \texttt{MUXAR} is shorthand for the right hand side of the rewrite, where the \texttt{SUM} represents array reduction:
\begin{align*} %\label{eqn:mux_array}
\lsub{t}(\lsub{s}(\lsub{q}a &\times \lsub{r}b) + \lsub{s}(\lsub{q}c \times \lsub{r}(\sim(\lsub{r}b)))) \rightarrow \\
\lsub{t}(\texttt{SUM}(&\lsub{s}(\lsub{q}(\lsub{1}b[0] ? \lsub{q}a : \lsub{q}c) << \lsub{u}0), \\
& \lsub{s}(\lsub{q}(\lsub{1}b[1] ? \lsub{q}a : \lsub{q}c)<<\lsub{u}1),..., \\
& \lsub{s}(\lsub{q}(\lsub{1}b[r-1] ? \lsub{q}a : \lsub{q}c)<<\lsub{u}(r-1)))).
\end{align*}
Logic synthesis is capable of exploiting this optimization if it identifies an expression of the form $(a\times b) + (c\times \sim b)$, but we must indicate the merging opportunity to our tool.

We added the ``Constant Expansion'' rules to re-express multiplication of a variable by a constant. These rules allow us to recreate results from the literature described in Section \ref{subsec:datapath_opt} on the MCM problem.
In addition to the explicit rewrite rules, constant folding is implemented as an e-class analysis in 
\texttt{egg}~\cite{Willsey2021Egg:Saturation}.

\subsection{Extraction} \label{subsect:extraction}
Having applied rewrites to the e-graph until saturation or timeout limits are reached, we now must extract the optimal design from potentially infinitely many choices. The extraction process must select a set of e-classes to implement and for each e-class, which e-node within that class to implement, subject to the constraint that the e-class children of each selected e-node must also be selected. Choosing an optimal design requires some metric that allows us to discriminate between competing implementations. Industrial circuit design is typically judged on area, latency and power consumption. In this contribution we will only use an area metric, therefore our definition of optimal will be the smallest circuit implementation. %Area is a simple to calculate but fundamental metric since we cannot turn to pipelining, like we can for challenging delay constraints.

We have developed a theoretical area estimate cost function in terms of the number of two-input gates required for the operator. It assigns a cost per operator that is a function of the input and output bitwidths. Table \ref{tab:operator_arch} lists the architectures on which we base the cost of the more complex operators. We introduce different costs for when at least one of the operands is a constant. The cost of a complete design is then the sum of the operator costs, and the objective is to minimise this cost. 

The major benefit of using a theoretical cost metric as opposed to using metrics derived from logic synthesis or HLS tools is the computation speed which, when combined with equality saturation, enables effective design space exploration. Cost metric validation is addressed in Section \ref{sect:cost_validation}.

By explicitly introducing rules for operator merging and sharing, we are able to define a cost for each node based only on its type and argument bitwidths, capturing downstream synthesis optimizations for \texttt{SUM}, \texttt{MUXAR} and \texttt{FMA}.
%The theoretical cost function is local, meaning the cost of a node is only a function of the node itself and its children. Such local cost estimates of operators within a circuit design typically fail to capture operator merging and sharing, two crucial features of hardware design. Introduction of the \texttt{SUM}, \texttt{MUXAR} and \texttt{FMA} operators and the associated ``Merging Ops'' rewrite rules directly captures the post-synthesis cost implications of merging operators.

When extracting an RTL implementation from an e-graph, one is immediately faced with the question of common subexpressions. Common subexpressions are naturally extracted as part of the e-graph construction process \cite{Willsey2021Egg:Saturation} and ideally we would want to utilize this information in the resulting hardware, for example extracting $(x+1) \times (x+1)$ as $\texttt{let } y = x+1\text{ in }y \times y$. However, this makes the extraction problem an inherently global problem over the e-classes, in the sense that the optimal e-node implementation for a given e-class may depend on the selected e-node implementation of the other e-classes in the graph. Previous solutions have solved this by posing optimal extraction as an integer linear programming (ILP) problem \cite{Smith2021PurePearl,Wang2020SPORES:Algebra}, and we follow the same approach in this work. 

Using the notation defined in Section \ref{subsec:equality_saturation}, for each e-node $n \in \mathcal{N}$ we associate a cost, $\text{cost}(n)$, given by the theoretical cost function, and a binary variable $x_n \in \{0,1\}$, which indicates whether $n$ is implemented in the final extracted RTL. Our objective is to minimize the total implementation cost, as described by (\ref{eqn: min_cost}).  (\ref{eqn: choose_child}) then guarantees that for every node $n$, we implement a node from each of its child e-classes. Lastly we introduce $\mathcal{S}$, the set of e-classes representing the desired expressions to implement in RTL. (\ref{eqn: all_outputs}) then ensures that all these outputs are produced by the final RTL.

\begin{align}
    \text{minimize: } \sum_{n \in \mathcal{N}} \text{cost}(n)x_n \text{ subject to:} \label{eqn: min_cost}\\
    \forall (n,c) \in E. \; x_n \leq \sum_{n' \in \mathcal{N}_c} x_{n'} \label{eqn: choose_child}\\
    \forall c \in {\mathcal S}. \; \sum_{n \in \mathcal{N}_c} x_n = 1.\label{eqn: all_outputs}
\end{align}
E-graphs may contain cycles, {\em e.g.} $x + 0 \to x$ induces a cycle. By introducing a topological sorting variable, $t_c$ for each e-class $c$, and associated constraints (\ref{eqn:topo_sort}) where $N$ is the number of e-classes and $\mathcal{C}(n)$ is the e-class containing node $n$, we ensure that the output expression is acyclic.
%This formulation works for acyclic e-graphs, however additional constraints are required if the e-graph contains cycles, as the final RTL extracted for a combinational logic design must itself be acyclic. We introduce a topological sorting variable for each e-class $c$, $t_c$, and impose the following constraint, where $N$ is the number of e-classes and $\mathcal{C}(n)$ is the e-class containing node $n$.
\begin{equation} \label{eqn:topo_sort}
    \forall (n,k) \in E\quad t_{\mathcal{C}(n)} - Nx_n - t_k \geq 1 - N
\end{equation}

If we select a node $n\in \mathcal{N}_c$ with child $k$, i.e. $x_n = 1$, this constraint simplifies to $t_c\geq t_k + 1$ to get a topologically sorted result, whereas in the case $x_n = 0$, the constraint is vacuously satisfied. We use the open source GLPK solver to calculate solutions to this ILP \cite{Makhorin2008GLPKKit}. 

We deploy the ILP extraction method in cases where sharing common subexpressions offers some improvement, otherwise for improved performance we can resort to the standard \texttt{egg} extraction 
method~\cite{Willsey2021Egg:Saturation}.

%%%%%%%%%%%%%%%%%%%%%%%%%%%%%%%%%%%%%%%%%%%%%%%%%%%%%%%%%%%
% RESULTS
%%%%%%%%%%%%%%%%%%%%%%%%%%%%%%%%%%%%%%%%%%%%%%%%%%%%%%%%%%%
\section{Results} \label{sect:results}
The RTL test cases are automatically optimized using our \texttt{egg}-based implementation and optimized RTL is extracted. Original and optimized RTLs are synthesized using Synopsys Design Compiler for a TSMC 7nm cell library. We proved the formal equivalence of the original and optimized RTLs using Synopsys HECTOR technology, a formal equivalence checking tool that runs in minutes on these testcases.

The original and optimized RTLs were synthesised at the minimum delay of the slowest of the two, and then similarly at the minimum area of the larger of the two designs. These represent comparisons towards the endpoints of a standard area-delay curve for a design. Table \ref{tab:results_table} summarises the results. In Figure \ref{fig:f_smooth_area_delay}, we present the area-delay profiles for the competing architectures of a Smoothing Kernel, which highlights the points of comparison used in Table \ref{tab:results_table}.

We consider two sets of examples. First we demonstrate how the tool exploits complex datapath blocks to optimize designs. Then we consider bitwidth-dependent architectures, where the optimal design may vary as a function of bitwidth.

\begin{table*}[t]
    \centering
    \caption{Logic synthesis results using Synopsys Design Compiler. We compared the automatically optimized results at the minimum delay which both designs could meet and at the smallest area that both designs could meet.}
    \begin{tabular}
    {|m{.2\textwidth}
     |r r r
     |r r r
     |}
        \hline
        Benchmark & \multicolumn{3}{c|}{Min Achievable Delay} & \multicolumn{3}{c|}{Min Achievable Area}  \\
                 & Delay (ns) & Orig Area ($\mu m^2$) & Opt Area ($\mu m^2$) & Area ($\mu m^2$) & Orig Delay (ns) & Opt Delay (ns)\\
        \hline
        Smoothing Kernel                           & 0.289 & 550  & 158 \textbf{(\texttt{-}71\%)} & 150  & 1.25 & 0.29 \textbf{(\texttt{-}77\%)}\\
        FIR Filter Kernel                          & 0.611 & 1710 & 679 \textbf{(\texttt{-}60\%)} & 570  & 1.38 & 2.48 \textbf{(\texttt{+}80\%)}\\
        ADPCM Decoder \cite{Lee1997MediaBench:Systems} & 0.102 & 103  & 102 \textbf{(\texttt{-} 1\%)} & 32   & 0.72 & 0.45 \textbf{(\texttt{-}38\%)}\\ 
        Shifted FMA                                & 0.181 & 310  & 210 \textbf{(\texttt{-}32\%)} & 81   & 0.97 & 0.57 \textbf{(\texttt{-}41\%)}\\
        MCM Solution                               & 0.132 & 90   & 104 \textbf{(\texttt{+}15\%)} & 40   & 0.72 & 0.56 \textbf{(\texttt{-}22\%)}\\
        \hline        
    \end{tabular}

    \label{tab:results_table}
\end{table*}

\begin{table}
    \centering
    \caption{Benchmark complexity and resulting e-graphs size after rewriting. $\dagger$ means that we removed the associativity of addition rewrite to limit e-graph growth. ILP column indicates whether ILP or standard \texttt{egg} extraction was used.}
    \begin{tabular}{|c|r|r|c|r|}
        \hline
        Benchmark         & Ops & E-graph Nodes & ILP & Runtime (sec) \\
        \hline
        Smoothing Kernel              & 17 & 27,000 & No & 140\\
        FIR Filter Kernel $\dagger$   & 23 & 550    & Yes & 100\\
        ADPCM Decoder                 & 9  & 6,700  & No & 19 \\ 
        Shifted FMA                   & 3  & 22     & No & 0.04 \\
        MCM Solution                  & 3  & 4,900  & Yes & 31 \\
        \hline
    \end{tabular}
    \label{tab:benchmark+ops}
\end{table}

\subsection{Datapath Optimizations using Complex Blocks} \label{subsect: results_datapaths}
\begin{figure}
    \centering
    \includegraphics[width=\columnwidth]{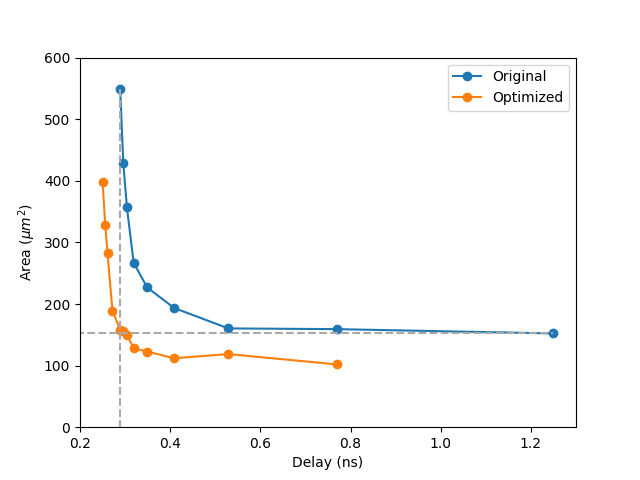}
    \caption{Area-delay profile of the Smoothing Kernel. The horizontal/vertical grey lines represents the minimum area/delay comparison points.}
    \label{fig:f_smooth_area_delay}
\end{figure}

The first benchmark is a kernel from a media module, which is an industrially relevant example supplied by Intel. It was manually optimized by a single engineer over the course of one week. Our tool automatically matches the results obtained via manual optimization, making use of the ``Merge Mult Array'' rewrite. Figure \ref{fig:f_smooth_area_delay} shows the area-delay curves for the original and optimized architectures. The optimized design performs strictly better, reducing the minimum achievable delay by 13\% with a 28\% area reduction.

Next we have a set of benchmarks taken directly from~\cite{dataflow2008verma}, which are intended to show how this prior work is generalized by our optimization tool. The first example is a finite-impulse response (FIR) filter with 8 taps. The second example, a computational kernel of  the ADPCM decoder \cite{Lee1997MediaBench:Systems}, is an approximation to a 16$\times$4 multiplier. These two examples are optimized by deploying the ``Arithmetic Logic Exchange'' class of rewrites described in Section \ref{subsect:rewrites}, to cluster additions together. The FIR filter example is particularly interesting as it also introduces a multiple constant multiplication (MCM) problem~\cite{Hartley1996SubexpressionMultipliers}. Since the ILP formulation of extraction correctly accounts for the cost of common subexpressions, the minimal area solution it extracts maximally shares common subexpressions, to generate $\{2\times x, 3\times x,...,7\times x\}$. For example, $2\times x$ and $3 \times x$ are constructed as follows
\begin{equation}
    x_2= x<<1, \hspace{0.5em} x_3= x_2 + x. \label{eqn:fir_mcm_0}
\end{equation}

\begin{figure}
    \centering
    \includegraphics[width=0.35\columnwidth]{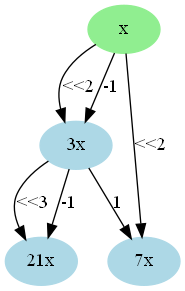}
    \caption{Data-flow graph representing an optimal adder tree to compute the set $3\times x, 7\times x$ and $21\times x$. The blue nodes represent additions which generate the result in the node label.}
    \label{fig:mcm_example_soln}
\end{figure}

Looking more generally at the MCM problem, using the ``Constant Expansion'' rewrites, we are able to match the operator count of a solution from \cite{Gustafsson2007AProblems}. An example from this paper generates adder graphs to compute $\{3 \times x, 7\times x, 21\times x\}$. The optimal design from our tool maximally shares common subexpressions to generate the data-flow graph shown in Figure \ref{fig:mcm_example_soln}, which uses 3 addition/subtraction operations to generate the results. Since logic synthesis likely implements constant multiplication using a CSD representation \cite{Ercegovac2004DigitalArithmetic}, sharing common subexpressions generates a $15\%$ larger circuit than the basic architecture for small delay targets as the design does not  match the performance of CSD.

Finally we consider a shifted FMA, which discovers the opportunity to merge a multiplication and addition that were originally separated by a left shift $(a\times b) << S + c$. Using the ``Left Shift Mult'' and ``FMA Merge'' rewrites this can be implemented as an FMA block. The approach proposed by Verma, Brisk and Ienne misses this opportunity since it will not move the shift over the multiplication \cite{dataflow2008verma}.

\subsection{Bitwidth Dependent Optimal Architecture} \label{subsect: bw_dept_arch}
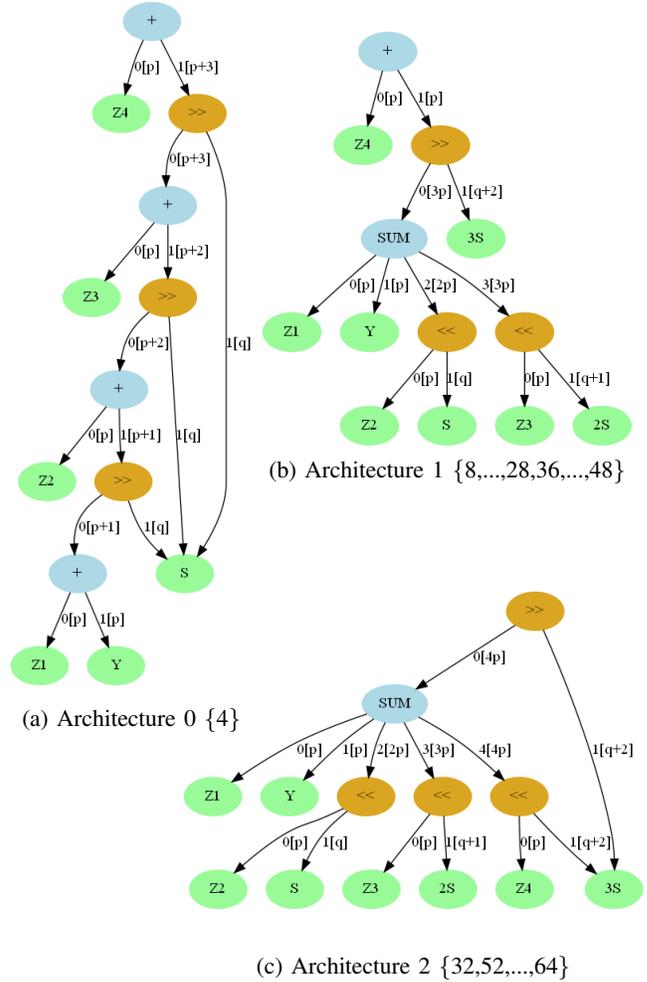
\begin{figure}
     \centering    
     \begin{tikzpicture}
        
        \node at (4.2, 1.4) {\resizebox{5cm}{!}{\includegraphics{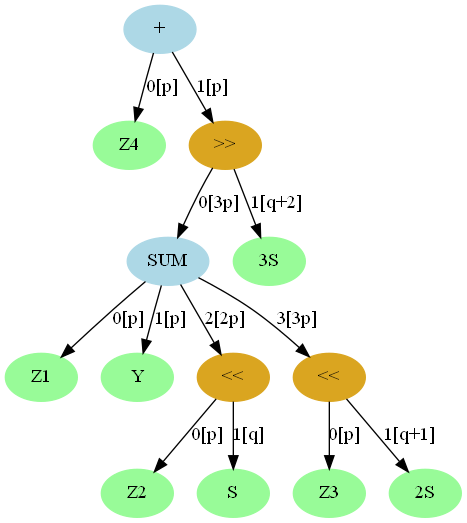}\label{fig:fir_4_arch_1}}};
        \node at (3.75, -5.4) {\resizebox{6.2cm}{!}{\includegraphics{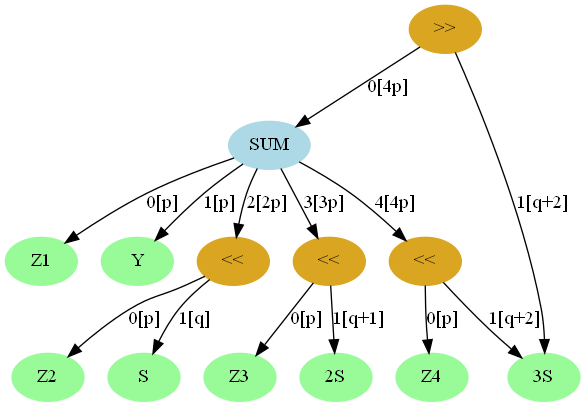}\label{fig:fir_4_arch_2}}};
        \node at (0,0) {\resizebox{3.3cm}{!}{\includegraphics{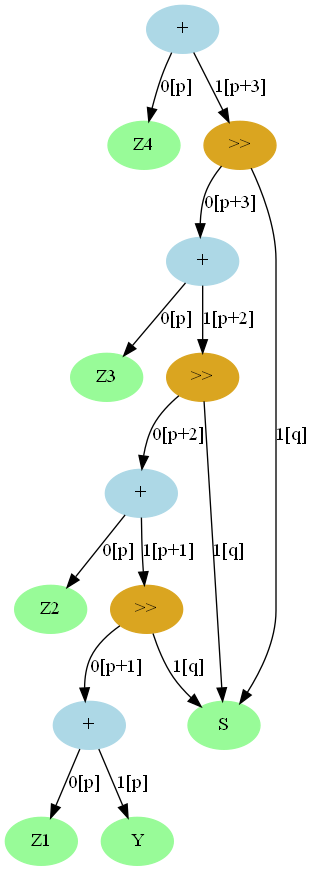}\label{fig:fir_4_arch_0}}};
        \node at (0, -5) {\small (a) Architecture 0 \{4\}};
        \node at (4.2, -1.7) {\small (b) Architecture 1 \{8,...,28,36,...,48\}};
        \node at (3.75, -8.3) {\small (c) Architecture 2 \{32,52,...,64\}};
  \end{tikzpicture}
     \caption{Simplified FIR filter data-flow graphs representing optimal architectures for different choices of the input bitwidth parameter $p$ and shift bitwidth parameter $q$. The sets in curly braces are bitwidths for which that architecture is optimal. In these graphs $Zi = Xi \times Ai$ and $2S$ and $3S$ are computed according to (\ref{eqn:fir_mcm_0}).}
     \label{fig: 4 tap fir}
 \end{figure}

The last set of results considers parameterizable RTL, as in general, RTL engineers do not generate alternative architectures for different bitwidths. As a result, designers may be sacrificing performance. To demonstrate this we again consider the FIR filter, but the 4 tap variant, where the original architecture is shown in Figure \ref{fig: 4 tap fir}, which we will refer to as Architecture 0. We optimized Architecture 0 using our tool for increasing bitwidths and observed that the selected architecture varied between Architectures 0, 1 and 2 (Figure~\ref{fig: 4 tap fir}). The difference between Architectures 1 and 2 is that the addition involving $Z4$ in Architecture 1 has been pushed back over the right shift, incurring the cost of an extra left shift, but saving a full carry-propagate adder. 

The architectural choices are determined by the theoretical two-input gate cost metric, discussed in Section \ref{subsect:extraction}. We synthesized each of the three architectures for bitwidths $4, 8,...64$, updating the delay target each time to the minimum delay that all three architectures could meet at that bitwidth. The theoretically optimal architecture for each bitwidth can be seen in Figure~\ref{fig: 4 tap fir}. For 56\% of the bitwidths, the architecture selected by the theoretical cost metric gave the lowest area result from logic synthesis. The ability to always choose the minimum area design according to logic synthesis using the theoretical metric is limited, since logic synthesis always incorporates delay considerations into its results and there is noise in the results.
%we do not currently consider delay in the model. 
We demonstrate this `noise floor' in Section \ref{sect:cost_validation}. 

%Figure \ref{fig:arch_choice_comparison} demonstrates that the optimal architecture chosen by our theoretical metric agrees with that chosen by logic synthesis in 56\% of the test cases. These results will be further discussed in Section \ref{sect:cost_validation} but it is clear that the theoretical model fails to capture the jumps in logic synthesis results seen for larger bitwidths. \gc{It's not clear to me as a reader what the implications of this are. Is it important? Does it kill the whole project? Why is he telling me this? Indeed, {\em is} it important? If not, maybe we should just scrap this paragraph.}

\subsection{Performance}
For these test cases only the ``Shifted FMA'' e-graph saturated as we limited the exploration to 10 iterations of e-graph rewriting \cite{Willsey2021Egg:Saturation}. For test cases using ILP extraction we limited the solver to a 100 second time budget, which means that the FIR Filter solution is classified as feasible but is possibly sub-optimal. There is a tradeoff between growing the e-graph and solving the ILP efficiently. Only the MCM solution included the ``Constant Expansion'' rewrite class since these rewrites lead to exponential growth of the e-graph. 

%%%%%%%%%%%%%%%%%%%%%%%%%%%%%%%%%%%%%%%%%%%%%%%%%%%%%%%%%%%
% COST METRIC VALIDATION
%%%%%%%%%%%%%%%%%%%%%%%%%%%%%%%%%%%%%%%%%%%%%%%%%%%%%%%%%%%
\section{Cost Metric Validation} \label{sect:cost_validation}
This section evaluates the theoretical cost metric in terms of its ability to steer optimal architecture choices. To do so we compare the theoretical cost to logic synthesis area costs. The results from Table \ref{tab:results_table} demonstrate that for all test cases, the theoretically chosen architecture improved at least the performance or the area of the design.

When determining the accuracy of a cost estimate, it is necessary to consider inherent variability of the logic synthesis process. Small non-functional tweaks, {\em e.g.} changing a variable name in RTL code, can have 
impact on the synthesis results. This forms a `noise floor' against which any theoretical cost model can be validated. To evaluate this noise floor, we used a technique known as fuzzing \cite{Miller1990AnUtilities}, which involves automatically generating random mutations to a program. We fuzz the RTL allowing two types of 
semantics-preserving mutations: variable renaming and swapping the order of always/assign blocks \cite{Thomas2008TheLanguage} in the code, modifications which one would not expect to have a meaningful impact on synthesis results.

\begin{figure}
    \centering
    \includegraphics[width=.95\columnwidth]{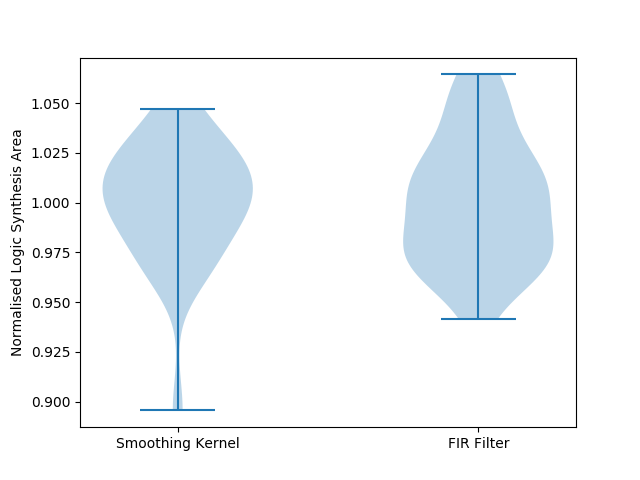}
    \caption{A violin plot depicting the logic synthesis area results for 30 fuzzed designs of the Smoothing Kernel and the FIR Filter at a 0.5ns delay target. For each violin, the area results are normalised by the mean.}
    \label{fig:fuzzing_violin}
\end{figure}

We provide results for the Smoothing Kernel and 4 tap FIR Filter, synthesizing 30 fuzzed designs in each case at relevant delay targets. Variability of the results is shown in Figure \ref{fig:fuzzing_violin}. 

\begin{figure}
    \centering
    \includegraphics[width=.95\columnwidth]{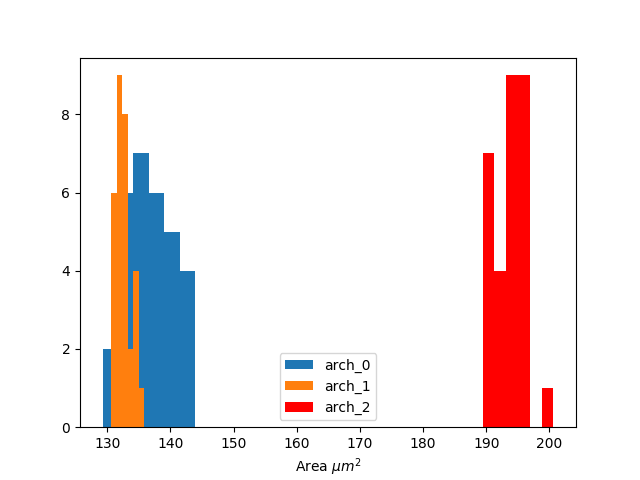}
    \caption{Histogram plot of logic synthesis area results for 30 fuzzed designs for each of the three FIR Filter architectures for 12 bits (Fig \ref{fig: 4 tap fir}).}
    \label{fig:histogram_12_bit}
\end{figure}

Figure \ref{fig:histogram_12_bit} highlights how this noise can affect the architectural choices made in Section \ref{subsect: bw_dept_arch}. For 12 bit inputs the synthesis results for fuzzed Architectures 0 and 1 overlap, with Architecture 1 generating lower area results on average whilst Architecture 0 obtains the minimum area result. This noise is not captured by the theoretical cost metric, as these fuzzed designs are theoretically identical. Applying this to other bitwidth inputs there are cases where there is clearly an optimal choice. 

\section{Conclusion}
This paper has demonstrated the application of e-graphs and equality saturation to the RTL datapath optimization problem. Applying a precisely defined set of bitwidth dependent equivalence preserving transformations, in the form of rewrites, we efficiently explore the design space and extract optimized RTL using \texttt{egg}. We also quantify the noise floor in logic synthesis results to understand the limits of a theoretical cost metric.

The results show that this automated rewriting technique can match the results of a skilled hardware engineer within a short timescale. The tool was able to achieve up to 71\% area improvement and up to 77\% delay improvement. We also demonstrate that automatic RTL optimization can generate different architectures for different bitwidth designs, since the tradeoff points are bitwidth dependent.

%%%%%%%%%%%%%%%%%%%%%%%%%%%%%%%%%%%%%%%%%%%%%%%%%%%%%%%%%%%
% FUTURE WORK
%%%%%%%%%%%%%%%%%%%%%%%%%%%%%%%%%%%%%%%%%%%%%%%%%%%%%%%%%%%
Future work will address the limitations of an area-only cost metric by incorporating delay and power allowing us to generate Pareto optimal solution curves \cite{Ustun2022IMpress:HLS}. We will expand the domain to tackle floating point operations and incorporate greater support for automated design verification. We will also address the scalability limits of applying equality saturation by incorporating intelligent design space search procedures. 

\section*{Acknowledgment}
The authors would like to thank Yann Herklotz for his assistance in setting up the fuzzing experiments. 
%\gcc{We would also like to thank Intel Corporation for their support and funding for this research.}{} \gc{It's atypical for a paper coauthored by an industry person to thank their employer, but can reinstate it if useful!} 



\bibliographystyle{IEEEtran}
\bibliography{references.bib}
%

\end{document}

%% file: flow_diagram.tex
\begin{tikzpicture}

% Boxes
\node [shape=rectangle,draw = black, text width = 1.2cm, text centered] at (0,0) (input) {Input Verilog};
\node [shape=rectangle,draw = black, minimum width=1.5cm, minimum height=1cm] at (2,0) (egraph) {E-graph};
\node [shape=rectangle,draw = black] at (2,-1.5) (rewrites) {Rewrite};
\node [shape=rectangle,draw = black, text width = 1.3cm, text centered] at (4,0) (extract) {Extract (ILP)};
\node [shape=rectangle,draw = red, dashed, minimum width=2cm, minimum height=3cm] at (2,-0.5) (egg) {};
\node [shape=rectangle,draw = black, text width = 1.5cm, text centered] at (6,0) (output) {Optimised Verilog};
\node [text=red] at (1.35,0.8) (egg_text) {\texttt{egg}};

\draw [->,very thick] (input) edge (egraph);
\path [->,very thick] (rewrites.east) edge[bend right] (egraph.south east);
\draw [-,very thick] (egraph.south west) edge[bend right] (rewrites.west);

\draw [->,very thick] (egraph) edge (extract);
\draw [->,very thick] (extract) edge (output);
\end{tikzpicture}

%% file: main.bbl
\begin{thebibliography}{10}
\providecommand{\url}[1]{#1}
\csname url@samestyle\endcsname
\providecommand{\newblock}{\relax}
\providecommand{\bibinfo}[2]{#2}
\providecommand{\BIBentrySTDinterwordspacing}{\spaceskip=0pt\relax}
\providecommand{\BIBentryALTinterwordstretchfactor}{4}
\providecommand{\BIBentryALTinterwordspacing}{\spaceskip=\fontdimen2\font plus
\BIBentryALTinterwordstretchfactor\fontdimen3\font minus
  \fontdimen4\font\relax}
\providecommand{\BIBforeignlanguage}[2]{{%
\expandafter\ifx\csname l@#1\endcsname\relax
\typeout{** WARNING: IEEEtran.bst: No hyphenation pattern has been}%
\typeout{** loaded for the language `#1'. Using the pattern for}%
\typeout{** the default language instead.}%
\else
\language=\csname l@#1\endcsname
\fi
#2}}
\providecommand{\BIBdecl}{\relax}
\BIBdecl

\bibitem{Synopsys2019CodingSynthesis}
{Synopsys}, ``{Coding Guidelines for Datapath Synthesis},'' Synopsys, Mountain
  View, Tech. Rep., 12 2019.

\bibitem{Synopsys2021DesignS-2021.06-SP2}
------, ``{Design Compiler User Guide S-2021.06-SP2},'' Synopsys, Mountain
  View, Tech. Rep., 6 2021.

\bibitem{Verma2009ChallengesCircuits}
A.~K. Verma, P.~Brisk, and P.~Ienne, ``{Challenges in automatic optimization of
  arithmetic circuits},'' in \emph{Proceedings - IEEE Symposium on Computer
  Arithmetic}, 2009, pp. 213--218.

\bibitem{dataflow2008verma}
------, ``{Data-flow transformations to maximize the use of carry-save
  representation in arithmetic circuits},'' \emph{IEEE Transactions on
  Computer-Aided Design of Integrated Circuits and Systems}, vol.~27, no.~10,
  pp. 1761--1774, 2008.

\bibitem{Willsey2021Egg:Saturation}
M.~Willsey, C.~Nandi, Y.~R. Wang, O.~Flatt, Z.~Tatlock, and P.~Panchekha,
  ``{Egg: Fast and extensible equality saturation},'' \emph{Proceedings of the
  ACM on Programming Languages}, vol.~5, no. POPL, 2021.

\bibitem{Gustafsson2007AProblems}
O.~Gustafsson, ``{A difference based adder graph heuristic for multiple
  constant multiplication problems},'' in \emph{IEEE International Symposium on
  Circuits and Systems}, 2007, pp. 1097--1100.

\bibitem{Hartley1996SubexpressionMultipliers}
R.~I. Hartley, ``{Subexpression Sharing in Filters Using Canonic Signed Digit
  Multipliers},'' \emph{IEEE Transactions on Circuits and Systems}, vol.~11,
  1996.

\bibitem{Ercegovac2004DigitalArithmetic}
M.~D. Ercegovac and T.~Lang, \emph{{Digital arithmetic}}.\hskip 1em plus 0.5em
  minus 0.4em\relax Elsevier, 2004.

\bibitem{Nelson1980TechniquesVerification}
C.~G. Nelson, ``{Techniques for program verification},'' Ph.D. dissertation,
  Stanford University, 1980.

\bibitem{Joshi2002Denali:Superoptimizer}
R.~Joshi, G.~Nelson, and K.~Randall, ``{Denali: A goal-directed
  superoptimizer},'' in \emph{Proceedings of the ACM SIGPLAN Conference on
  Programming Language Design and Implementation (PLDI)}, 2002.

\bibitem{Tate2009EqualityOptimization}
R.~Tate, M.~Stepp, Z.~Tatlock, and S.~Lerner, ``{Equality saturation: A new
  approach to optimization},'' in \emph{ACM SIGPLAN Notices}, vol.~44, no.~1,
  2009.

\bibitem{Panchekha2015AutomaticallyExpressions}
P.~Panchekha, A.~Sanchez-Stern, J.~R. Wilcox, and Z.~Tatlock, ``{Automatically
  improving accuracy for floating point expressions},'' \emph{ACM SIGPLAN
  Notices}, vol.~50, no.~6, pp. 1--11, 2015.

\bibitem{Smith2021PurePearl}
G.~H. Smith, A.~Liu, S.~Lyubomirsky, S.~Davidson, J.~McMahan, M.~Taylor,
  L.~Ceze, and Z.~Tatlock, ``{Pure tensor program rewriting via access patterns
  (representation pearl)},'' in \emph{Proceedings of the 5th ACM SIGPLAN
  International Symposium on Machine Programming}, 2021.

\bibitem{Wang2020SPORES:Algebra}
Y.~R. Wang, S.~Hutchison, J.~Leang, B.~Howe, and D.~Suciu, ``{SPORES:
  Sum-product optimization via relational equality saturation for large scale
  linear algebra},'' \emph{Proceedings of the VLDB Endowment}, vol.~13, no.~11,
  2020.

\bibitem{Wolf2013Yosys-ASuite}
C.~Wolf and J.~Glaser, ``{Yosys-A Free Verilog Synthesis Suite},'' in
  \emph{Proceedings of Austrochip}, 2013.

\bibitem{Steele1990CommonLanguage}
G.~Steele, \emph{{Common LISP: the language}}.\hskip 1em plus 0.5em minus
  0.4em\relax Elsevier, 1990.

\bibitem{Beaumont-Smith2001ParallelDesign}
A.~Beaumont-Smith and C.-C. Lim, ``{Parallel prefix adder design},'' in
  \emph{Proceedings - IEEE Symposium on Computer Arithmetic}, 2001, pp.
  218--225.

\bibitem{Koren2018ComputerAlgorithms}
I.~Koren, \emph{{Computer arithmetic algorithms}}.\hskip 1em plus 0.5em minus
  0.4em\relax AK Peters/CRC Press, 2018.

\bibitem{Thomas2008TheLanguage}
D.~Thomas and P.~Moorby, \emph{{The Verilog{\textregistered} hardware
  description language}}.\hskip 1em plus 0.5em minus 0.4em\relax Springer
  Science {\&} Business Media, 2008.

\bibitem{Barrett2016TheSMT-LIB}
C.~Barrett, P.~Fontaine, and C.~Tinelli, ``{The Satisfiability Modulo Theories
  Library (SMT-LIB)},'' www.SMT-LIB.org, 2016.

\bibitem{Makhorin2008GLPKKit}
\BIBentryALTinterwordspacing
A.~Makhorin, ``{GLPK (GNU linear programming kit)},'' 2008. [Online].
  Available: \url{http://www.gnu.org/s/glpk/glpk.html}
\BIBentrySTDinterwordspacing

\bibitem{Lee1997MediaBench:Systems}
C.~Lee, M.~Potkonjak, and W.~H. Mangione-Smith, ``{MediaBench: A tool for
  evaluating and synthesizing multimedia and communications systems},'' in
  \emph{Proceedings of the Annual International Symposium on
  Microarchitecture}, 1997.

\bibitem{Miller1990AnUtilities}
B.~P. Miller, L.~Fredriksen, and B.~So, ``{An empirical study of the
  reliability of UNIX utilities},'' \emph{Communications of the ACM}, vol.~33,
  no.~12, pp. 32--44, 1990.

\bibitem{Ustun2022IMpress:HLS}
E.~Ustun, I.~San, J.~Yin, C.~Yu, and Z.~Zhang, ``{IMpress: Large Integer
  Multiplication Expression Rewriting for FPGA HLS},'' in \emph{2022 IEEE 30th
  Annual International Symposium on Field-Programmable Custom Computing
  Machines (FCCM)}, 2022, pp. 1--10.

\end{thebibliography}
